\documentclass[12pt]{article}
\usepackage{graphicx}
\usepackage{amsmath}
\usepackage{amssymb}

\newcommand\pubdate{\today}
\def\D0bar{\overline D{}^0}
\def\K0bar{\overline K{}^0}
\def\DDbar{D^0 - \overline D{}^0}
\newcommand{\beq}{\begin{equation}}
\newcommand{\eeq}{\end{equation}}
\newcommand{\bey}{\begin{eqnarray}}
\newcommand{\eey}{\end{eqnarray}}
\newcommand{\bdm}{\begin{displaymath}}
\newcommand{\edm}{\end{displaymath}}
\newcommand{\nn}{\nonumber}

\def\eeqn{\end{equation}}

\def\beqa{\begin{eqnarray}}
\def\eeqa#1{\label{#1}\end{eqnarray}}
\def\eeqan{\end{eqnarray}}

\let\bar=\overbar

\def\Dslash{\not{\hbox{\kern-4pt $D$}}}
\def\dslash{\not{\hbox{\kern-2pt $\del$}}}

\def\msb{{\bar{\ssstyle M \kern -1pt S}}}

\def\Title#1{\begin{center} {\Large #1 } \end{center}}
\def\Author#1{\begin{center}{ \sc #1} \end{center}}
\def\Address#1{\begin{center}{ \it #1} \end{center}}

\newcommand\pubblock{\rightline{\begin{tabular}{l} 
         \pubdate  \end{tabular}}}
\newenvironment{Abstract}{\begin{quotation}  }{\end{quotation}}
\newenvironment{Presented}{\begin{quotation} \begin{center}
             PRESENTED AT\end{center}\bigskip
      \begin{center}\begin{large}}{\end{large}\end{center} \end{quotation}}
\def\Acknowledgements{\bigskip  \bigskip \begin{center} \begin{large}
             \bf ACKNOWLEDGEMENTS \end{large}\end{center}}

\begin{document}
\begin{titlepage}
\pubblock

\vfill
\Title{Dominant $1/m_c$ Contribution To The Mass Difference in $D^0$-${\overline D}^0$ Mixing}
\vfill
\Author{Gagik Yeghiyan}
\Address{Department of Physics, Grand Valley State University, Allendale, MI 49401, USA}
\vfill
\begin{Abstract}
We re-visit the problem of non-perturbative contribution to the mass difference in $\DDbar$ mixing within the Standard Model (SM).
There are two known approaches to take this contribution into account. In the long-distance approach one analyzes contribution to the
$\DDbar$ from exclusive channels due to production of intermediate charmless hadronic states. In the short-distance approach one employs the Operator Product Expansion (OPE) techniques with the non-perturbative contribution coming from the diagrams containing low energy-momentum quark states - quark-antiquark condensates. Within the latter approach, the non-perturbative contribution to the normalized mass difference in $\DDbar$ mixing, $x_D = \Delta M_D / \Gamma_D$, is estimated to be $few \times 10^{-3}$, that is to say very close to the experimental value of $x_D$. We attempt to verify quantitatively this estimate here. We find that the actual prediction of the short-distance approach is few times less
than the estimate above and order of magnitude less than the experimental value of $x_D$.
\end{Abstract}
\vfill
\begin{Presented}
The Fifth International Workshop On Charm Physics\\
Honolulu, Hawaii,  May 13--17, 2012
\end{Presented}
\vfill
\end{titlepage}
\def\thefootnote{\fnsymbol{footnote}}
\setcounter{footnote}{0}
%

$\DDbar$ oscillations may serve as an efficient tool for indirect searches for New Physics (NP).
Recent discovery of CP-violation in the charm quark involving processes \cite{4,5} with the experimental data
possibly exceeding the SM predictions
has revived the interest to the charm sector as the place where a signal for New Physics (NP) may be detected.

Within the perturbative approach, the Standard Model contribution
to the mass and the width difference in $\DDbar$ mixing is suppressed by several orders of magnitude \cite{36,6} as compared
to the experimental values of these quantities. The
experimental values of the normalized mass and width differences in $\DDbar$ mixing \cite{3},
\bey
x_D^{exp} = \frac{\Delta M_D^{exp}}{\Gamma_D} = (6.3 \pm 1.9) \times 10^{-3}  \label{1} \\
y_D^{exp} = \frac{\Delta \Gamma_D^{exp}}{2 \Gamma_D} = (7.5 \pm 1.2) \times 10^{-3} \label{2}
\eey
may be either due to non-perturbative effects within the Standard Model \cite{2,15,7,8,9} or due to New Physics effects
(see \cite{23,basic,37} and references therein). Whether
one of these two effects dominates or (\ref{1}) and/or (\ref{2}) are due to interference of these two effects,
is still unclear mainly because of the lack of precise evaluation of the non-perturbative SM contribution.

In this paper the problem of non-perturbative contribution to the mass difference in $\DDbar$ mixing within the
Standard Model is readdressed.

Two ways of taking into account the non-perturbative contribution to $\DDbar$ mixing are presently known. The
long-distance approach considers the exclusive channels of $\DDbar$ mixing, when this process occurs due to production of
intermediate charmless hadronic states. It has been estimated using this approach that $x_D^{SM} \sim 10^{-2}$ and
$y_D^{SM} ~ \sim 10^{-2}$.

Another (short-distance) approach is based on the use of Operator Product Expansion (OPE) techniques with the non-perturbative
effects coming from the diagrams containing low energy-momentum quark states - quark-antiquark condensates.
The contribution of these diagrams corresponds to higher-order (as compared to the dimension-six 
four-fermion operators) 
$1/m_c$ terms in the OPE. The estimates done
using this approach predict \cite{7} $x_D^{SM} \sim few \times 10^{-3}$, and $y_D \sim 10^{-3}$.
Apparent discrepancy between the results of the two approaches may be both because of the Operator Product Expansion failure (as
$\mu_{had}/ m_c$ is not a small quantity), and because of possible cancelations in the contribution of different exclusive channels.

In this paper we attempt to evaluate quantitatively the non-perturbative contribution to the mass difference in $\DDbar$ mixing
using the OPE techniques\footnote{Quantitative evaluation of subleading diagrams for the width difference has been
done in \cite{1}.}. Our preliminary results show that the actual prediction of the short-distance approach is few times less
than the estimate of ref.~\cite{7} and order of magnitude less than the experimental value of $x_D$.

Within the Standard Model, $\DDbar$ oscillations are analyzed using the two generation mixing approach (neglecting
the contribution of the diagrams with b-quark intermediate states - this approach is well justified (at least when the mass difference is
of the interest) due to the suppressed mixing of the third generation of quarks with the first two). With this assumption, $\DDbar$
mixing vanishes in the well-defined theoretical limit of exact flavor $SU(3)$ symmetry. In the real world, where the flavor $SU(3)$
is broken, the SM contribution to the $\DDbar$ is of course non-vanishing. Yet, it is suppressed by a certain power of the
strange-to-charm mass ratio. In particular, to the lowest order in the perturbation theory, $x_D^{LO} \propto m_s^4/m_c^4$
and $y_D^{LO} \propto m_s^6/m_c^6$. More precisely, for the mass difference \cite{6},
\bey
\nn
&&(x_D)^{LO}=
\frac{G_F^2 m_c^2}{3 \pi^2} \frac{f_D^2 M_D}{\Gamma_D}
\left(\frac{m_s^4}{m_c^4} \right)
\sin^2 \theta_C \cos^2 \theta_C \\
&& \times \left\{\frac{5}{4} \left[C_2^2 - 2 C_1 C_2 - 3 C_1^2 \right] \overline{B}_D^{(S)}
- C_2^2 B_D
\right\} \approx 1.7 \times 10^{-6} \label{3}
\eey
Thus, within the SM, to the lowest order in the perturbation theory $x_D$ is smaller by several orders of magnitude than its experimental value.

\begin{figure}[t]
\centering
\includegraphics{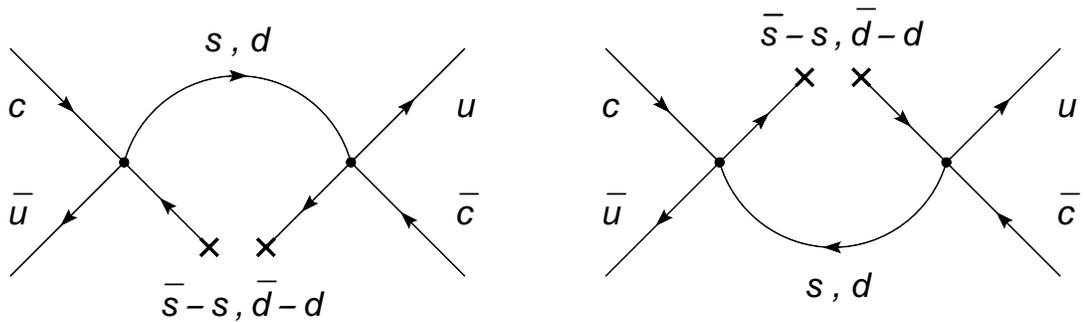}
\caption{Diagrams with two hanging quark lines.} \label{f2}
\end{figure}

It has been argued however \cite{2,15,7} that the SM contribution to $x_D$ may be significantly greater due to the diagrams involving
interactions of the valence charm and up quarks and antiquarks with (non-perturbative) low energy-momentum intermediate QCD background
quark states (light quark-antiquark condensates). The simplest case with a single low
energy-momentum quark state and the energy being transferred from $D^0$ to $\D0bar$ (and back) by a single light quark propagator is depicted in
Figure~\ref{f2}. The non-perturbative quark-antiquark
state is denoted by a pair of hanging light quark lines.

\begin{figure}[t]
\centering
\includegraphics{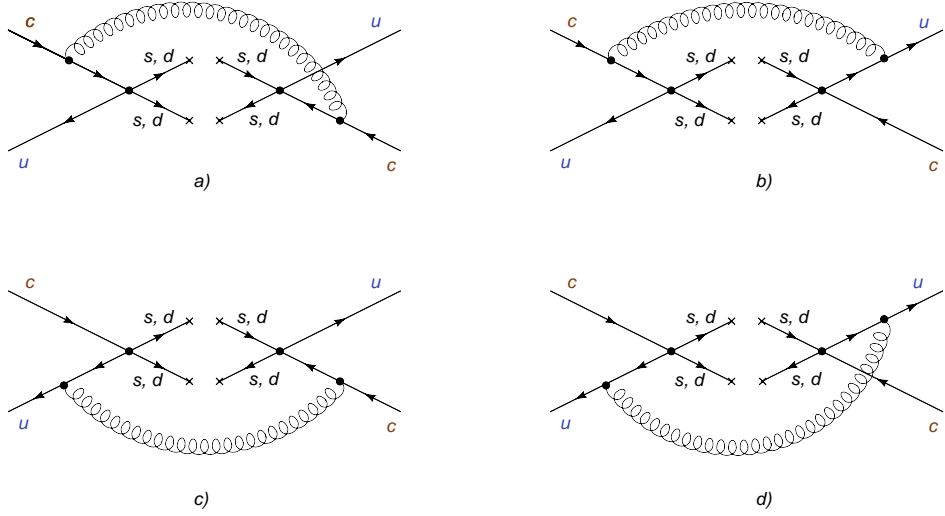}
\caption{Examples of diagrams with four hanging quark lines.} \label{f1}
\end{figure}

It is also possible that all the intermediate light quark states are low energy-momentum states and the energy from $D^0$ to $\D0bar$ and back
is transferred by a hard gluon. Examples of the diagrams with four hanging light quark lines (that describe such a scenario) are presented in Figure~\ref{f1}.

Note that each time one cuts a quark line, the strange or down quark propagator in the matrix element of $\DDbar$ transition must be
replaced by a dimension-three product of strange or down quark operators. Then an extra factor of $m_c^3$ must appear in the denominator
to preserve the dimensionality of the matrix element (and that of the mass and the width difference).
Thus, the diagrams in Fig.~\ref{f2} yield the matrix elements of
$d = 9$ local operator products or $1/m_c^3$ suppressed
terms in Operator Product Expansion.  Subsequently, the diagrams in Fig.~\ref{f1} yield the matrix elements of
$d = 12$ local operator products or $1/m_c^6$ suppressed
terms in Operator Product Expansion.

As the ratio of $\mu_{had}/m_c$ is not a small quantity, both the $1/m_c^3$ and the $1/m_c^6$ terms in OPE may be non-negligible or even
dominant provided that their suppression in powers of $m_s/m_c$ is weaker than that of the lowest-order term. It has been proven \cite{2}
using a group-theoretical analysis that mass insertion in each light quark propagator produces a factor of $m_s^2$, whereas
the interaction of the valence charm and up quarks and antiquarks with a QCD background quark-antiquark state produces a factor of $m_s$ only. Thus, one may expect a softer
flavor $SU(3)$ suppression in the diagrams with hanging quarks!

The rule of the suppression power counting in the diagrams with hanging quarks states as follows \cite{7}:
\begin{itemize}
\item{Cutting a quark line, we pay the price of a power suppression  $\sim \mu_{had}^3/m_c^3$. Yet, flavor $SU(3)$ suppression (in this fermion line)  is
$m_s/\mu_{had}$ only and there is no loop factor. Altogether  we have the enhancement $\sim 4 \pi^2 \mu_{had}^2 /(m_s m_c)$,
 which may yield $(x_D)^{d=9}  \sim 10^{-4}$.}
 \item{Subsequently, cutting two quark lines (and inserting a gluon propagator that transfers the energy from $D^0$ to $\D0bar$ and back), produces an enhancement factor of  $\sim 4 \pi^2  (4 \pi \alpha_s) \mu_{had}^4 /(m_s^2 m_c^2)$ (with $\alpha_s = \alpha_s (m_c)$).
     This may yield $(x_D)^{d=12} \sim  few \times 10^{-3}$.}
 \end{itemize}

 The latter estimate however turns to be somewhat altered by some additional suppression factors that emerge in the quantitative evaluation
 of the diagrams with hanging quarks. We illustrate this on the subset of the diagrams in Fig.~\ref{f1} where the gluon couples to the external
 (charm and up) quark lines. The more comprehensive evaluation of all relevant diagram with hanging quarks will be presented in our
 forthcoming publication \cite{10}.

For the diagrams in Fig.~\ref{f1} one gets
\bey
\nn
(x_D)_{external}^{d=12}=
\left(\frac{2 G_F^2 m_c^2}{9} \right)
(4 \pi \alpha_s) \frac{f_D^2 M_D}{\Gamma_D}
\left(\frac{m_s^2}{m_c^2} \right) \left(\frac{\Lambda^4}{m_c^4} \right)
\sin^2 \theta_C \cos^2 \theta_C \\
\times \left\{\frac{4 C_1}{3} \left[3 C_1 + 2 C_2 \right] B_D -
 \left[C_2^2 + \frac{8 C_1 C_2}{3} + 4 C_1^2 \right] \overline{B}_D^{(S)} \right\}\label{4}
\eey
where $\Lambda$ is a non-perturbative parameter of the order of hadronization scale or smaller
(or $\Lambda \sim 1$~GeV or smaller) that is related to the light quark-antiquark condensates.
It is defined as
\bey
m_s^2 \Lambda^4 = \langle 0 | \Biggl[ \left(\bar{s}(0) \gamma_\mu P_L s(0) - \bar{d}(0) \gamma_\mu P_L d(0) \right)
\left(\bar{s}(0) \gamma^\mu P_L s(0)
- \bar{d}(0) \gamma^\mu P_L d(0) \right) \label{5} \\ \nn
- 2 \bar{s}(0) \gamma_\mu P_L d(0) \bar{d}(0) \gamma^\mu P_L s(0) \Biggr] | 0 \rangle
\eey

Choosing $\Lambda = 1 GeV$, one gets
\beq
(x_D)_{external}^{d=12} = -0.76 \times 10^{-3} \label{6}
\eeq
While the negative sign may be flipped when evaluating the remaining diagrams with hanging quarks, it is apparent that
our result is somewhat smaller than the estimate discussed above. This is mainly due to the fact that the loop factor
$1/(3 \pi^2)$ present in Eq.~(\ref{3}) does not just disappear (as was assumed in ref.~\cite{7}) but is
replaced by another suppression factor, $2/9$, due to Fierz re-arrangements of the relevant color indices
when deriving (\ref{4}). More detailed analysis of the power $1/N_{color}$ suppression effects is going to be
presented in \cite{10}.

Thus, our quantitative evaluation of the non-perturbative contribution to the mass difference in $\DDbar$ mixing
yields the value of $x_D$ that is few times smaller than the made estimates and by order of magnitude less than the
experimental output for this quantity. The interpretation of our result is rather ambiguous. As mentioned above, it may be due to
the failure of Operator Product Expansion (as $\mu_{had}/m_c$ is not a small quantity). Another issue may be
the validity of the factorization approach used to derive (\ref{4}). On the other hand, it is also possible that the
SM contribution to $x_D$ is indeed small (provided strong cancelations in the contribution of the exclusive channels
occur), and the experimental value of $x_D$ is due to New Physics effects.

In conclusion we attempted to evaluate quantitatively the non-perturbative contribution to the mass difference in $\DDbar$ mixing
within the Standard Model
using the OPE techniques. Our preliminary results show that within this approach the SM prediction for $x_D$ is few times less
than the previously made estimate and order of magnitude less than the experimental value of this quantity.

\Acknowledgements
Author is grateful to the organizers of Charm-2012 workshop at University of Hawaii and the Center of Scholarly and Creative Excellence at Grand Valley State University for financial support, without which this presentation would not be possible. \\
This work has been partially supported by the grant DOE~DE-FGO2-96ER41005.

\bigskip 

\begin{thebibliography}{99} 

\bibitem{4} T. Aaltonen et al. (CDF Collaboration), Phys. Rev. D 85, 012009 (2012).
[arXiv:1111.5023 [hep-ex]].
\bibitem{5} R. Aaij et al. (LHCb Collaboration), Phys. Rev. Lett. 108, 111602 (2012).
[arXiv:1112.0938 [hep-ex]].

\bibitem{36}    A.~Datta and D.~Kumbhakar,
   Z. Phys.  {\bf C 27}, 515 (1985).

\bibitem{6} E. Golowich and A. A. Petrov, Phys. Lett. B 625, 53 (2005).

\bibitem{3} Heavy Flavor Averaging Group (HFAG), http://www.slac.stanford.edu/xorg/hfag/

\bibitem{2} H. Georgi, Phys. Lett. B 297, 353 (1992).

\bibitem{15} T. Ohl, G. Ricciardi, E. H. Simmons, Nucl. Phys. B 403, 605, (1993). 

\bibitem{7} I.Bigi, N. Uraltsev, Nucl. Phys. B 592, 92 (2001).

\bibitem{8} A. F. Falk, Y. Grossman, Z. Ligeti, Y. Nir and A. A. Petrov, Phys.Rev. D 69, 114021 (2004).

\bibitem{9} A. F. Falk, Y. Grossman, Z. Ligeti and A. A. Petrov, Phys. Rev. D 65, 054034 (2002).

\bibitem{23}
E.~Golowich, J.~Hewett, S.~Pakvasa and A.~A.~Petrov,
  Phys.\ Rev.\  D {\bf 76}, 095009 (2007).

\bibitem{basic} A. A. Petrov, G. K. Yeghiyan, Phys. Rev. D 77, 034018 (2008).

\bibitem{37} G. K. Yeghiyan, Phys. Rev. {\bf D 76}, 117701 (2007).

\bibitem{1} M. Bobrowski, A. Lenz, T. Rauh,  arXiv:1208.6438 [hep-ph]; \\
M. Bobrowski, A. Lenz, 	arXiv:1009.4545 [hep-ph], 	PoS ICHEP2010, 193 (2010).

\bibitem{10} E. Golowich, A. A. Petrov and G. Yeghiyan, to be published.

\end{thebibliography}

\end{document}